\documentclass[%
twocolumn,
superscriptaddress,
 amsmath,amssymb,
 aps,
]{revtex4-2}

\usepackage[english]{babel} 
\usepackage[T1]{fontenc}
\usepackage[ansinew]{inputenc}

\usepackage{amsmath}
\usepackage{amsthm}
\usepackage{amsfonts}
\usepackage{bbold}
\usepackage{xcolor}
\usepackage{soul}
\usepackage{cancel}

\usepackage{graphicx}
\usepackage{dcolumn}
\usepackage{bm}
\newcommand{\B}{\mathbf{B}}

\begin{document}

\preprint{v004 - \today}

\title{Magnetic flux threading a planar holed superconductor}
\author{Jaume Cunill-Subiranas}
\thanks{These two authors contributed equally to this work.}
\affiliation{Departament de F\'isica, Universitat Aut\`onoma de Barcelona, 08193 Bellaterra, Barcelona, Spain}
\author{Natanael Bort-Soldevila}
\thanks{These two authors contributed equally to this work.}
\affiliation{Departament de F\'isica, Universitat Aut\`onoma de Barcelona, 08193 Bellaterra, Barcelona, Spain}
\author{Nuria Del-Valle}
\affiliation{Departament de F\'isica, Universitat Aut\`onoma de Barcelona, 08193 Bellaterra, Barcelona, Spain}
\author{Carles Navau}
\email{carles.navau@uab.cat}
\thanks{Serra H\'unter Professor}
\affiliation{Departament de F\'isica, Universitat Aut\`onoma de Barcelona, 08193 Bellaterra, Barcelona, Spain}


\begin{abstract}
We analytically evaluate the currents in a planar-holed ideal (perfect magnetic shielding) superconductor when an axial dipole is located along the system's axis, obtaining a simple relation between the fluxoid (in this geometry, it equals the flux) threading the hole, the net current circling it, and the magnitude and position of the dipole.  This expression has to be added to the basic screening equations describing the superconductor to understand the different physical situations that can appear. Thus, several scenarios are studied: field and zero-field cooling cases, where the flux threading the hole is fixed; superconductors with a slit where the net current circling the hole is zero; and self-inductance case, where the flux threading the hole is only due to the screening currents. All the presented results are analytical for the axial dipole and are given as a function of the distance of the dipole from the plane of the superconductor. The results obtained can serve as a benchmark for recent and ongoing micromechanical experiments aimed at developing ultrasensitive gravimeters or applications searching for exotic spin-dependent interactions and dark matter, for example.
\end{abstract}


\maketitle

\section{Introduction}
\label{sec:intro}
The levitation of a magnetic dipole over a planar superconductor (SC) is an especially paradigmatic and simple case to study the interaction of magnetic elements with superconductors. It is well known that if the superconductor is ideal (perfectly in the Meissner state) and infinite, the force the dipole receives is that of a dipole image  \cite{Wei1996}. If the superconductor is a type-II superconductor, some flux lines can enter in the form of fluxoids and the frozen-image model can also serve to evaluate the interaction of such a superconductor with a dipole \cite{Kordyuk1998}. In general, the interaction of dipoles (or small permanent magnets) with superconductors has been profusely studied, especially in the field of magnetic levitation \cite{Lin2006,Navau2013,Bernstein2020,Werfel2012}. Recently, the levitation of ferromagnets over superconductors has received renewed interest, especially at the micron-scale, as they can be used as high-sensitivity inertial sensors, magnetometers or gravimeters, provided they can be isolated from the environment \cite{Wang2019,Timberlake2019}. In Ref.  \cite{Timberlake2019}, for example, a 4 mg magnet was levitated over a lead superconducting disk at 300 mK and at a background helium pressure of 10$^{-10}$ mbar, which could achieve sensitivities of about 3$\cdot$10$^{-15}$ g/$\sqrt{\rm Hz}$ with ideal conditions and vibration isolation. In \cite{Fuchs2024}, a submillimeter magnetic particle levitated on a superconducting trap was gravitationally coupled to a kilogram source mass approximately half a meter away. These experiments, and others \cite{Ahrens2024,Vinante2020}, demonstrate the importance of developing the theory for the levitation of magnets over superconductors.

\begin{figure}[b]
    \centering    
  \includegraphics[width=0.48\textwidth]{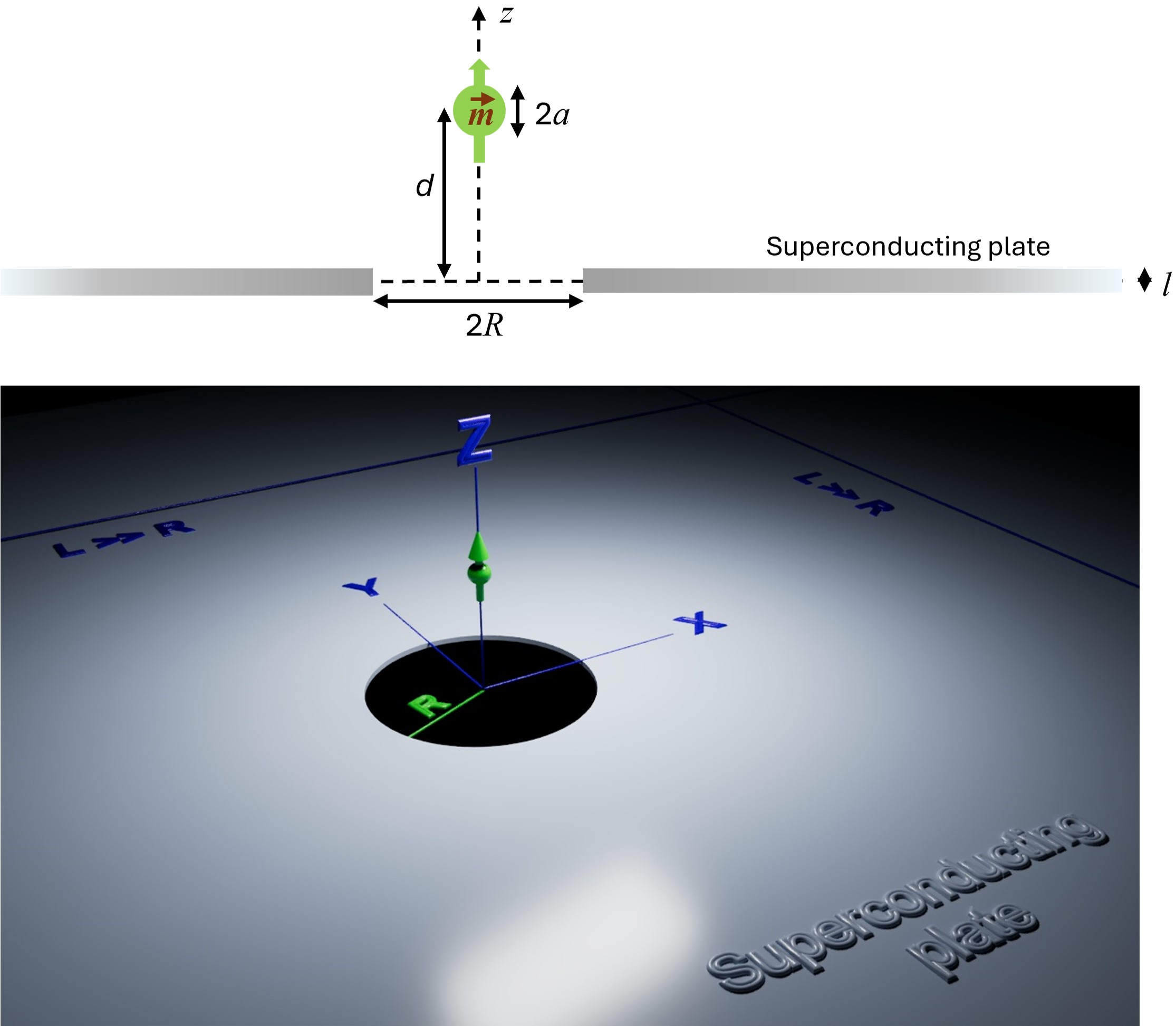}    \caption{Sketch (2D and 3D views) of the studied system. The plane is considered wide (size $L\gg R$), and an ideal superconductor (perfect screening). The dipole is assumed to point in the $z-$direction.}
    \label{fig:sketch}
\end{figure}

Most of the models used previously consider the superconductor in the Meissner state and assume simplified models for the frozen-flux in field-cooled conditions.
The consideration of non-simply-connected geometries in the superconductor adds extra complexity and potential to the levitation systems. The fluxoid threading a closed line $\mathcal{L}$ lying inside the superconductor, defined as the flux of magnetic induction field ${\bf B}$ threading the surface bounded by $\mathcal{L}$, plus the line integral of the supercurrent equation along the curve $\mathcal{L}$ must be quantized and fixed, in static situations. This property has been exploited in sensing with the use of superconducting quantum interference devices (SQUIDs)  \cite{Einzel2011,Clarke2000} or, to cite some recent applications, to search invisible axion dark matter \cite{Du2018,Higgins2024} or in proposals for quantum magnetomechanics experiments \cite{RomeroIsart2012,Cirio2012,Navau2021,Latorre2023,BortSoldevila2024}.

In addition to this fluxoid conservation, the induced currents must arrange themselves to make ${\bf B}=0$ in the interior of the superconductor. 
However, these conditions do not suffice to fully determine the currents in the superconductor and the field they create. As shown in Refs. \cite{Babaei2003,Brandt2004} there are several situations with different solutions: (i) there is some magnetic flux threading the hole, with or without external field; (ii) the superconductor has been cooled in a zero-applied field and the fluxoid crossing the hole remains zero; or (iii) there is an externally applied field but the net current circulating the hole is zero, for example. In the system we shall describe (a large planar superconductor with a hole), the external field will be given by a finite source (i.e., a dipole). Thus, provided that the superconducting plane is large enough, and using a line $\mathcal{L}$ along the exterior border of the superconductor, it is safe to consider that the currents at $\mathcal{L}$ are zero and the fluxoid reduces to the flux of magnetic induction field ${\bf B}$ threading the hole.

In this work, inspired by magnetomechanical systems for creating inertial and force sensors \cite{Prat-Camps2017,Gieseler2020,Vinante2022,Fuwa2023,Fuwa2023_b}, where a planar superconductor with a hole is used as a magnetic levitation trap, we aim to study a non-simply-connected geometry of the superconductor coupled with a magnetic dipole. 
The goal is to obtain simple enough equations to describe this system so that it could serve as the simplest benchmarking case and as the starting point for other more complex theoretical, numerical, and experimental studies. We plan to give an analytical description of the currents induced in the superconductor and the total field around it and, as important as this, to identify and quantify in a simple relation the relevant magnitudes that allow us to discriminate between the physical scenarios that can appear depending on the magnetic history of the superconductor. We shall focus on the different physical situations that emerge from considering non-simply-connected superconductors. A sketch of the studied system can be seen in Fig. \ref{fig:sketch}.

The paper is structured as follows. In Sec. \ref{sec:description} we describe the system and comment in detail on the assumptions done, and how the relevant magnitudes should be evaluated from a general axially symmetric applied field. In Sec. \ref{sec:AxialDipole} we present the analytical solutions for the current densities and magnetic field, as well as the central flux-current-field equation relating the total flux threading the hole, the net current circling it, and the applied dipolar field. In Sec. \ref{sec:scenarios} we discuss how the different physical scenarios emerge from understanding the flux-current-field equation and how the particular cases are obtained from the general ones. Finally, we present our conclusions in Sec. \ref{sec:conclusion}.

\section{General equations}
\label{sec:description}

Consider a thin planar superconductor, very wide (infinite) in the $x$ and $y$ dimensions, occupying $z\in(-l/2,l/2)$. Consider that there is a circular hole of radius $R\gg l$. We consider that the superconductor follows the London equation (without vortices), but assume that the London penetration depth $\lambda$ obeys $\lambda<l/2$ or, if $\lambda>l/2$, that the two-dimensional screening length $\Lambda=2\lambda^2/l$ obeys $\Lambda\ll R$, such that the boundary condition is that the normal component of the magnetic induction is zero on the surface of the superconductor (perfect screening) \cite{Babaei2003}. Through this work, we use standard cylindrical coordinates $(\rho,\varphi,z)$ with the origin on the centre of the hole. 

All the above assumptions allow us to consider \cite{Clem1994} that there is a two-dimensional current density in the superconductor ${\bf K}(\rho,\varphi)=\int_{-l/2}^{l/2}{{\bf J}(\rho,\varphi,z)\,{\rm d}z}$, (${\bf J}$ is the volume current density) distributed in such a way that the normal component of the total magnetic induction  ${\bf B}$ (due to the currents and to an eventual external applied field) is zero on the surface of the superconductor, which is located at $z=0$ (perfect shielding case). 
 
 The goal is to find the surface current distribution in the superconductor and the magnetic fields in the surrounding region. 

If there is an external field, we consider that it has cylindrical symmetry around the $Z$ axis and that its sources are localized outside the superconductor. In this case, the problem has cylindrical symmetry and the magnitudes do not depend on $\varphi$. Outside the superconductor, the vector potential has only $\varphi$-component and obeys the Laplace equation. Its general solution under these conditions is 
\begin{eqnarray}
  A_\varphi(\rho,z) = \int_{0}^{\infty}  C(k) J_1(k\rho) e^{-k|z|} \, {\rm d}k + A^{\rm a}_\varphi(\rho,z), \nonumber \\
	\mbox{   if } z\neq 0 \mbox{ or } \rho<R,
\label{eq:Aint}
\end{eqnarray}
where $C(k)$ is a function that should be determined by the boundary conditions, $J_n()$ are the Bessel functions of order $n$, and $A^{\rm a}_\varphi$ is the $\varphi$-component of the vector potential due to the external applied field. The integral term in the above equation is the potential vector due to the currents induced in the superconducting plane. The induction field, ${\bf B}=\nabla \times {\bf A}$, has $z$ and $\rho$ components. Outside the superconductor, they are given by
\begin{eqnarray}
  B_\rho(\rho,z) = \pm \int_{0}^{\infty}  C(k) k J_1(k\rho) e^{-k|z|} \,{\rm d}k + B^{\rm a}_\rho(\rho,z), \nonumber \\ \mbox{   if } z\neq 0 \mbox{ or } \rho<R,
\label{eq:Brhoint}\\
\label{eq:Bzint}B_z(\rho,z) = \int_{0}^{\infty} C(k) k J_0(k\rho) e^{-k|z|} \,{\rm d}k  + B^{\rm a}_z(\rho,z), \nonumber\\ \mbox{   if } z\neq 0 \mbox{ or } \rho<R. 
\end{eqnarray}
The $\pm$ sign can be substituted by $z/|z|$. $B^{\rm a}_\rho$ and $B^{\rm a}_z$ are the $\rho$ and $z$ components of the externally applied induction field, respectively.

The surface currents induced in the superconductor can be found from the discontinuity of the tangential component of ${\bf B}/\mu_0$ at the superconducting surface ($z=0$). This is,
\begin{equation}
  K_\varphi(\rho>R) = \frac{2}{\mu_0} \int_{0}^{\infty} C(k) k J_1(k\rho) \, {\rm d}k.
\label{eq:Kint}
\end{equation}

To obtain $C(k)$, one should use the boundary conditions at the $z=0$ plane.
 Indeed, for $z=0$ and $\rho<R$ there must be no discontinuity in $B_\rho$ (no currents there), and, for $z\rightarrow 0$ and $\rho>R$ (on the superconducting surface), the $z$-component of ${\bf B}$ has to be zero. Thus,
\begin{equation}
\begin{cases}
		\int_0^\infty {\rm d}k \,  \, C(k) \, k \, J_1(k\rho)  = 0,& \mbox{  if } \rho<R, \\  
		\int_0^\infty {\rm d}k \, C(k) \, k \, J_0(k\rho)  = - B^{\rm a}_z(\rho,0),& \mbox{  if } \rho>R.\label{eq:Csys} 
 \end{cases}
\end{equation}


For the sake of simplicity, we define the following dimensionless variables and functions:
\begin{eqnarray}
  t&=&k R, \label{eq:tnorm}\\
	x&=&\rho/R,\label{eq:xnorm}\\
	g(x)&=&B^{\rm a}_z(x R,0)/B_0, \\ \label{eq:gnorm}
	c(t)&=&- C(t/R) / (B_0 R^2). \label{eq.cnorm}
\end{eqnarray}
$B_0$ is a convenient normalization constant with units of induction field, that should be explicitly determined for the particular applied induction field $\B^{\rm a}$. 

Using the above definitions, Eq. \eqref{eq:Csys} becomes a dual integral equation for the normalized coefficients $c(t)$:
\begin{eqnarray} 
   \begin{cases}
			 \int_0^\infty \, {\rm d}t \, c(t) t J_1(tx) = 0, & \mbox{if } x<1, \\ 
		   \int_0^\infty \, {\rm d}t \, c(t) t J_0(tx) = g(x), & \mbox{if } x>1.    
		\label{eq:normdual}
		\end{cases}
\end{eqnarray}

It is important to note that there is no unique solution for the $c(t)$ function. Indeed, if $c(t)$ is a solution of Eq. \eqref{eq:normdual}, any other function $c'(t)=c(t)+c_0(t)$ such that 
\begin{eqnarray} 
  \begin{cases}
			 \int_0^\infty \, {\rm d}t \, c_0(t) t J_1(tx) = 0, & \mbox{if } x<1,  \\ 
		   \int_0^\infty \, {\rm d}t \, c_0(t) t J_0(tx) = 0, & \mbox{if } x>1,    
		\label{eq:gauge}
  \end{cases}
\end{eqnarray}
would also be solution of Eq. \eqref{eq:normdual}. The left-hand side of the above equations defines the Hankel transform of the first and zeroth order of the $c_0(t)$ function \cite{poularikas_book}. The function whose Hankel transform is zero in the given intervals of $x$ is $c_0(t)=F \sin(t)/t$, where $F$ is an arbitrary constant.

To find the solution of Eq. \eqref{eq:normdual}, we follow Ref. \cite{Mandal1988}. After reducing the system to a Fredholm integral equation of the second kind and using the Hankel transform to isolate the unknown function $c(t)$, one obtains:
\begin{eqnarray}
	  c(t)= t^{-1/2} \int_1^\infty r J_{1/2}(t r) \, g_2(r) \,{\rm d}r,
        \label{eq:csol}
\end{eqnarray}
where the function $g_2(r)$ is 
\begin{equation}
g_2(r)=-\sqrt{\frac{2}{\pi r}} \frac{\rm d}{{\rm d} r} \left[\int_r^\infty  (x^2-r^2)^{-1/2} x \, g(x) \,{\rm d}x \right].
\label{eq:g2sol}
\end{equation}
The current density on the superconducting surface ($x>1$) is obtained from Eq. \eqref{eq:Kint}, using Eq. \eqref{eq:csol}, and including the $c_0$ term:
\begin{equation}
    K_\varphi(x>1) = -\frac{2B_0}{\mu_0} \sqrt{\frac{2}{\pi}} \frac{1}{x} \int_1^x g_2(r) \frac{r^{3/2}}{\sqrt{x^2-r^2}} \,{\rm d}r + K_\varphi^F(x).
    \label{eq:Kphiwithg2}
\end{equation}
Similarly, from Eq. \eqref{eq:Bzint}, using Eq. \eqref{eq:csol}, we get the $z$ component of the $\B$ field at the hole ($x<1$, $z=0$):
\begin{eqnarray}
    B_z(x<1,0) &=& B_z^{\rm a} (x,0) -  B_0 \sqrt{\frac{2}{\pi}} \int_1^\infty g_2(r) \frac{r^{1/2}}{\sqrt{r^2-x^2}} \, {\rm d}r  \nonumber \\ &+& B_z^F(x).
    \label{eq:Bzwithg2}
\end{eqnarray}

In the above expressions, the terms $K_\varphi^F(x)$ and $B_z^F(x)$ come from the extra degree of freedom given by the conditions of Eq. \eqref{eq:gauge} with $c_0(t)=F \sin(t)/t$. The physical interpretation of the dimensionless constant $F$ will be given below. Then,
\begin{equation}
    K_\varphi^F(x>1)=-\frac{2B_0}{\mu_0} F \frac{1}{x\sqrt{x^2-1}}, \label{eq:KphiI}
\end{equation}
\begin{equation}
    B_z^F(x<1) =-B_0 F \frac{1}{\sqrt{1-x^2}}. \label{eq:BzI}
\end{equation}

The net current $I$ circling the hole, $I=\int_R^\infty K_\varphi(\rho) \, {\rm d} \rho $, can be found from Eq. \eqref{eq:Kint} as
\begin{equation}
    I  = -\frac{B_0 R \sqrt{2\pi}}{\mu_0}  \int_1^\infty r^{1/2} g_2(r) \, {\rm d}r - \frac{B_0 R \pi}{\mu_0} F .
    \label{eq:IVsF}
\end{equation}
The flux threading the hole, $\Phi = \int_0^R B_z(\rho,0) \, 2\pi \rho \, {\rm d}\rho$, can be found from Eq. \eqref{eq:Bzwithg2} as
\begin{eqnarray}
    \Phi &=& \Phi^{\rm a} \nonumber -  2 B_0 R^2 \sqrt{2\pi}\int_1^\infty g_2(r) r^{1/2}(r-\sqrt{r^2-1}) \, {\rm d}r  \nonumber \\ &-& 2\pi B_0 R^2 F,
    \label{eq:PhiVsF}
\end{eqnarray}
where $\Phi^{\rm a}$ is the flux threading the hole due exclusively to the externally applied field. 

Eqsuations \eqref{eq:IVsF} and \eqref{eq:PhiVsF} give the constant $F$ a physical meaning: $F$ is related to the total current circling the hole \textit{and} to the total flux threading it. Thus, if the net current circling the hole is given, the constant $F$ is no longer a free parameter but it is determined by Eq. \eqref{eq:IVsF} and the flux determined by Eq. \eqref{eq:PhiVsF}. On the contrary, if the flux is given, $F$ becomes determined by Eq. \eqref{eq:PhiVsF} and the net current by Eq. \eqref{eq:IVsF}. 

\section{Solution for an axial dipole}
\label{sec:AxialDipole}
The formulation of the previous section is valid as long as the externally applied field is cylindrically symmetric around the $Z$ axis. In this section, we will find explicit analytical relations for surface current densities, net currents, fields (in the $z=0$ plane), and fluxes, for the particular case of a dipolar axial applied field. 

Consider a small sphere (radius $a\ll R$) uniformly magnetized in the $z$ direction with magnetization $M$. This sphere is located on the $Z$ axis, at a position $z=d$.  It is known that the field generated by this sphere is uniform inside it and dipolar-like outside it \cite{jackson_book}. In particular, the $z$ component of the magnetic induction field at the $z=0$ plane is given by 
\begin{eqnarray}
  & &B_z^{\rm a}(\rho,0) = \nonumber \\  
	 &=& \begin{cases} 
			\frac{\mu_0 m}{4\pi}\frac{2}{a^3},& \mbox{if }  \rho^2+d^2 < a^2, \\
			\frac{\mu_0 m}{4\pi}\left(\frac{3d^2}{[\rho^2+d^2]^{5/2}}-\frac{1}{[\rho^2+d^2]^{3/2}}\right),& \mbox{if } \rho^2+d^2>a^2. 	
			\end{cases}	\nonumber \\	
\label{eq:Bzsph} 
\end{eqnarray}
In the above equation, $m=\frac{4}{3}\pi a^3 M$ is the magnetic moment of the equivalent dipole that can substitute the sphere when evaluating the field and vector potential outside the own sphere. The sphere is used here instead of a point-dipole to avoid infinities appearing when calculating the flux crossing the hole when $d=0$. The external field felt by the superconducting region is, thus, exactly dipolar-like.

In this case, a convenient choice of the $B_0$ constant is $B_0=\mu_0 m/(4\pi R^3)$. Defining $\delta = d/R$, and noting that $g(x)$ is only defined for $x>1$, one gets [Eqs.\eqref{eq:tnorm}-\eqref{eq:gnorm}]
\begin{equation}
g(x)=\frac{2\delta^2-x^2}{(\delta^2+x^2)^{5/2}}.
\end{equation}
The function $g_2(r)$ is, after Eq. \eqref{eq:g2sol},
\begin{equation}
    g_2(r) = -2\sqrt{\frac{2r}{\pi}}  \frac{\left(r^2-3\delta^2\right)}{\left(r^2+\delta^2\right)^3}.
\end{equation}

From Eqs. \eqref{eq:Kphiwithg2}-\eqref{eq:BzI}, and using the Eq. \eqref{eq:IVsF} to write the surface current density and the $z$ component of the induction field as a function of the net current circling the hole, one finds
\begin{widetext}
\begin{eqnarray}
 K_\varphi(x>1) &=& 
  \frac{2 I}{\pi  R x \sqrt{x^2-1}}
   +\frac{m}{\pi^2 R^3} \Bigg[ \frac{ \delta^2(1+\delta ^2)-\left(\delta ^2+2\right)
   x^2+x^4}{ \left(\delta ^2+1\right)  x \sqrt{x^2-1}
   \left(\delta ^2+x^2\right)^2} 
   -\frac{3 x \delta   }{ \left(\delta ^2+x^2\right)^{5/2}} \tan ^{-1}
   \left( \delta \sqrt{\frac{x^2-1}{\delta ^2+x^2}}
   \right) \Bigg],
   \label{eq:Kphidelta} 
\end{eqnarray}
\small
\begin{eqnarray}
     B_z(x<1,0) = B_z^{\rm a}(x,0) + \frac{\mu_0 I}{\pi  R \sqrt{1-x^2}} 
  +\frac{\mu_0 m }{2\pi^2 R^3} \Bigg[
   \frac{\delta ^2 \left(\delta ^2+2\right)-\left(2 \delta ^2+1\right)
   x^2}{\left(\delta ^2+1\right) \sqrt{1-x^2} \left(\delta ^2+x^2\right)^2} +\frac{ \left(x^2-2 \delta ^2\right) }{\left(\delta
   ^2+x^2\right)^{5/2}} \tan
   ^{-1}\left(\sqrt{\frac{\delta ^2+x^2}{1-x^2}}\right)  \Bigg]. \label{eq:Bzdelta}
\end{eqnarray}
\normalsize
\end{widetext}

The flux crossing the hole due to the applied field is 
\begin{equation}
    \Phi^{\rm a} = \frac{\mu_0 m}{2 R} \left(\frac{1}{ (1+\delta^2)^{3/2}}\right),
    \label{eq:phiappl}
\end{equation}
for all $\delta$, independently of the value of $a$. 

Evaluating the net current circling the hole and the flux threading it from Eqs. \eqref{eq:IVsF} and \eqref{eq:PhiVsF}, respectively, we obtain the following important flux-current-field relation (after eliminating the $F$ parameter):
\begin{equation}
\Phi=2\mu_0  R I + \frac{\mu_0 m }{\pi R} \left(\frac{1}{1+\delta^2}\right).
\label{eq:central}
\end{equation}
This is one of the central results of this work. This equation expresses, in a simple way, the relation between the applied field (through $m$ and $\delta$), the net current circling the hole ($I$), and the total flux threading it ($\Phi$). It says that one cannot independently set the flux and the circling current. At the same time, it indicates that the screening equations do not suffice to completely determine the currents in the superconductor. 

Equations \eqref{eq:Kphidelta}-\eqref{eq:Bzdelta} could be alternatively written in terms of the total flux $\Phi$ instead of $I$, using Eq. \eqref{eq:central}. The following section will show how different physical situations discriminate between possibilities. 

\section{Different physical scenarios}
\label{sec:scenarios}
\subsection{Field-cooling (FC) the superconductor with a centred dipole}
\label{sec:fc}
Consider the planar holed superconductor \textit{above} the critical temperature and the dipole (the small magnetized sphere) in the centre of the hole ($\delta=0$). The superconductor (now considered a non-magnetic material) does not influence the total field since it contains no currents. 
	
When cooling down the superconductor below the critical temperature, some currents must appear to make ${\bf B}=0$ in the interior of the superconductor (Meissner effect). Among all the possibilities of surface current density distribution that can satisfy this, the most energetically favourable is that with a circling net current equal to zero, since this condition does not add magnetic energy to the system (any net current $I$ circling the hole would add positive energy $L I^2/2$, being $L$ the self-inductance of the system). 

In this situation, the surface current density and the $z$ component of the induction field are, after Eqs. \eqref{eq:Kphidelta} and \eqref{eq:Bzdelta},
\begin{eqnarray}
K_{\varphi,{\rm FC_0}}(x>1)&=&\frac{m}{\pi^2 R^3} \left(\frac{x^2-2}{x^3 \sqrt{x^2-1}} \right),  \label{eq:KphiFC}\\
B_{z,{\rm FC_0}}(x<1,0) &=& B_z^{\rm a}(x,0)  \nonumber \\ &+& \frac{\mu_0 m}{2 \pi ^2 R^3} \Bigg[  -\frac{1}{x^2 \sqrt{1-x^2}} 
 \nonumber \\ &+&\frac{1}{x^3} \tan ^{-1}\left(\frac{x}{\sqrt{1-x^2}}\right) \Bigg].
\label{eq:BzFC}
\end{eqnarray}
These functions are plotted in Fig. \ref{fig:FC} (case $\delta=0$, black thick lines). Since the net current is zero, the surface current density changes in sign (at $x=\sqrt{2}$). According to Eq. \eqref{eq:central}, the flux threading the hole is \begin{equation}
\Phi_{{\rm FC_0}}=\frac{\mu_0 m}{\pi R}.
\label{eq:PhiFC}
\end{equation}
This flux can be compared with the flux that crosses a circle of radius $R$ in the $z=0$ plane due to the sphere alone, without the superconductor, $\frac{\mu_0 m}{2 R}$. It is seen that when cooling down the holed superconductor with the centred dipole the flux is reduced by a factor $\pi/2$. After emerging from the dipole, the magnetic field lines should bend toward the plane $z=0$. When approaching the $z=0$ plane, since the superconductor does not allow field lines to cross, some lines should go towards the hole but others toward infinity. This reduction factor contrasts with a finite ring with a uniform applied field perpendicular to the disk surface \cite{Babaei2003}, where there is some flux-focusing in the superconducting hole. The key point is that, in the present case, the external field is generated in the central region of the hole.

\subsection{Moving the dipole after field-cooling the superconductor}
Starting from the previous situation of a field-cooling superconductor with a centred dipole, the surface current distribution will change if the dipole is moved. The new surface current distribution should still yield ${\bf B}=0$ inside the superconductor. The flux threading the hole has to be kept (at $\Phi_{{\rm FC_0}}$) during the movement. There is no longer a need to have $I=0$ since the currents can take some energy from the external field change. In fact, the net current varies with the position of the dipole according to Eq. \eqref{eq:central}: 
\begin{equation}
I_{{\rm FC_0}}= \frac{m}{2\pi R^2} \left(\frac{\delta^2}{1+\delta^2}\right).
\end{equation}

\begin{figure}[tb]
	\centering
		\includegraphics[width=0.48\textwidth]{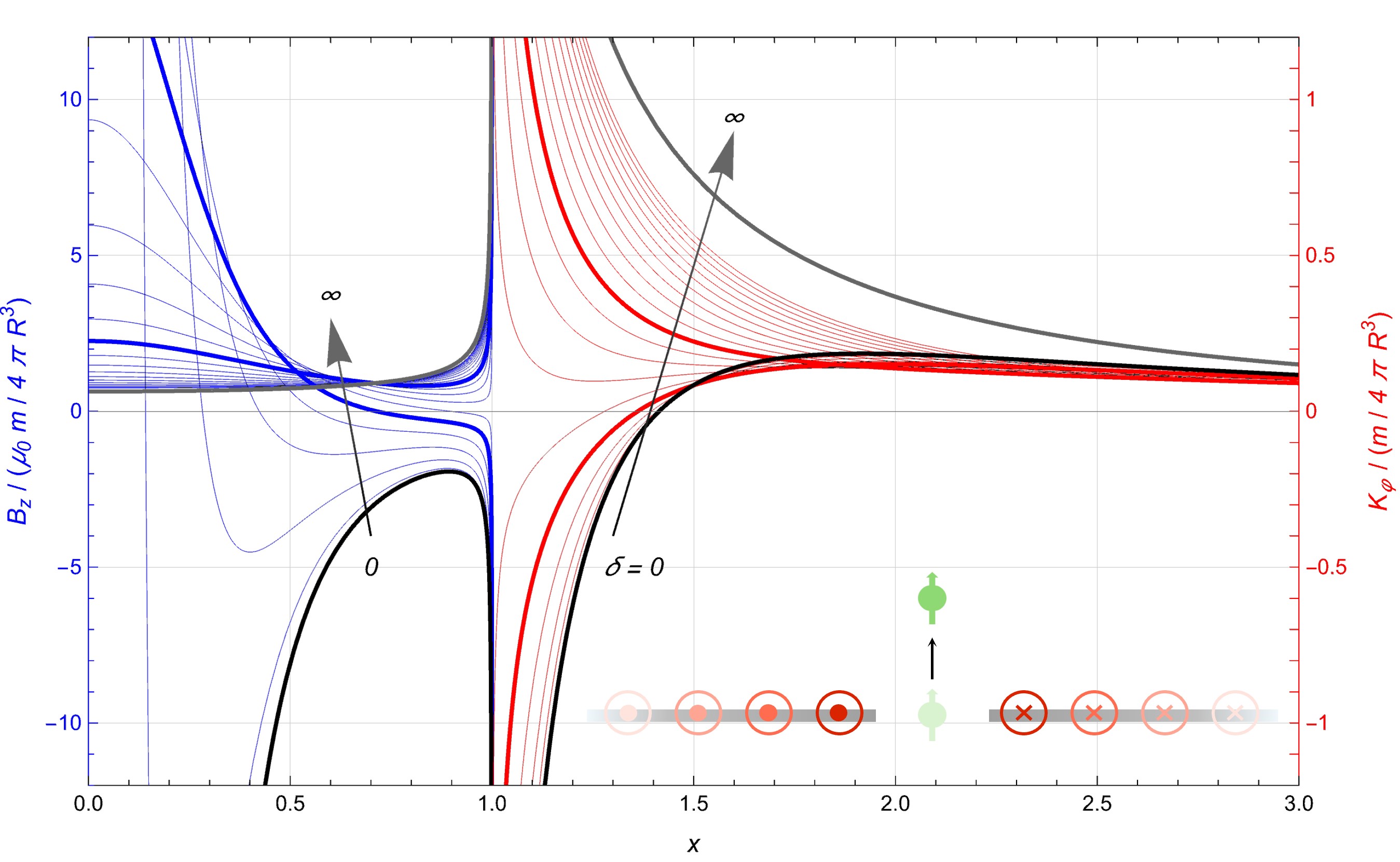}
	\caption{Current distribution $K_\varphi(x>1)$ and $B_z(x<1,0)$ component of the field at the hole when the superconductor has been field cooled with a dipole at the centre ($\delta=0$, highlighted in thick black line) and, after cooling, the dipole is moved along the $z-$axis. The plots are shown for $\delta=0$ to $\delta=2$ in intervals of $0.1$ following the arrows (the cases $\delta=0.5$ and $\delta=1$ are highlighted in thicker lines). The case $\delta\rightarrow \infty$ is also included and highlighted in a thick gray line. The inset indicates that the system has been FC with the dipole at center ($\delta=0$)  and moved towards $\delta\rightarrow \infty$. The sketched currents are those at $\delta\rightarrow \infty$.} 
	\label{fig:FC}
\end{figure}

The surface current density, as well as the $B_z$ component of the field in the $z=0$ plane, can be obtained from Eqs. \eqref{eq:Kphidelta} and \eqref{eq:Bzdelta}, substituting $I\rightarrow I_{{\rm FC_0}}$. They are plotted in Fig. \ref{fig:FC}, for different $\delta$'s. At large distances, $x\gg 1$, the surface current density decays as $\sim \rho^{-2}$, as in a Pearl vortex  \cite{Pearl1964}.

\subsection{Zero applied field after field cooling. Self-inductance.}

If the field is completely removed (if the sphere is moved away to infinity), only the circulating currents are responsible for the flux threading the hole. In this case, one can obtain an expression for the self-inductance of the system directly from Eq. \eqref{eq:central}, by setting $m=0$. One has $\Phi_L= 2\mu_0 I R$ and the self-inductance $L=\Phi_L/I$ is 
\begin{equation}
L=2\mu_0 R.
\label{eq:selfinductance}
\end{equation}
Alternatively, the self-inductance could be found from the current distributions and the field at the hole in the limit $\delta\rightarrow\infty$. In this case
\begin{eqnarray}
K_{\varphi,L}(x>1)&=&\frac{2 I}{\pi R} \left[\frac{1}{x\sqrt{x^2-1}}\right],
\label{eq:KA} \\
B_{z,L}(x<1,0)&=&
    \frac{\mu_0 I}{\pi R}\frac{1}{\sqrt{1-x^2}},
\end{eqnarray}
where the net current is $I=\Phi_{{\rm FC_0}}/(2\mu_0 R)$ if the initial flux was due to a central dipole, $\Phi_{{\rm FC_0}}$.  These expressions are plotted in Fig. \ref{fig:FC} with thick gray lines. 


\subsection{Field cooling with a given flux.}
A similar treatment can be done if the dipole is not at the centre when the superconductor is cooled below its critical temperature. If the dipole is at a given $\delta_{\rm FC}$ when the material transits to the superconducting state, the net current $I$ should be zero and the flux would be
\begin{equation}
\Phi_{\rm FC} = \frac{\mu_0 m}{\pi R}\left(\frac{1}{1+\delta_{\rm FC}^2}\right)
\end{equation}
after the transition. This flux should be maintained when the dipole moves (after the transition) and the net current given by Eq. \eqref{eq:central}. In all cases, the surface current densities and the field at the hole plane would be given by Eqs. \eqref{eq:Kphidelta} and \eqref{eq:Bzdelta}, with the adequate substitution of $I$ (which would depend on $\delta$).

\subsection{Zero-field cooling (ZFC)}
\label{sec:zfc}
Imagine now that the superconductor is cooled down in the absence of any applied field. There are no currents in the superconductor and no flux threading the hole:
\begin{equation}
\Phi_{\rm ZFC}=0.
\end{equation}
This flux is now maintained after variations of the external field. In our case, if we consider that the dipole is moved from infinity to a given position indicated by $\delta$, the net current that circles the hole is, from Eq. \eqref{eq:central},
\begin{equation}
I_{\rm ZFC}=- \frac{m}{2\pi R^2} \left(\frac{1}{1+\delta^2}\right).
\label{eq:Izfc}
\end{equation}

\begin{figure}[t!]
	\centering
		\includegraphics[width=0.48\textwidth]{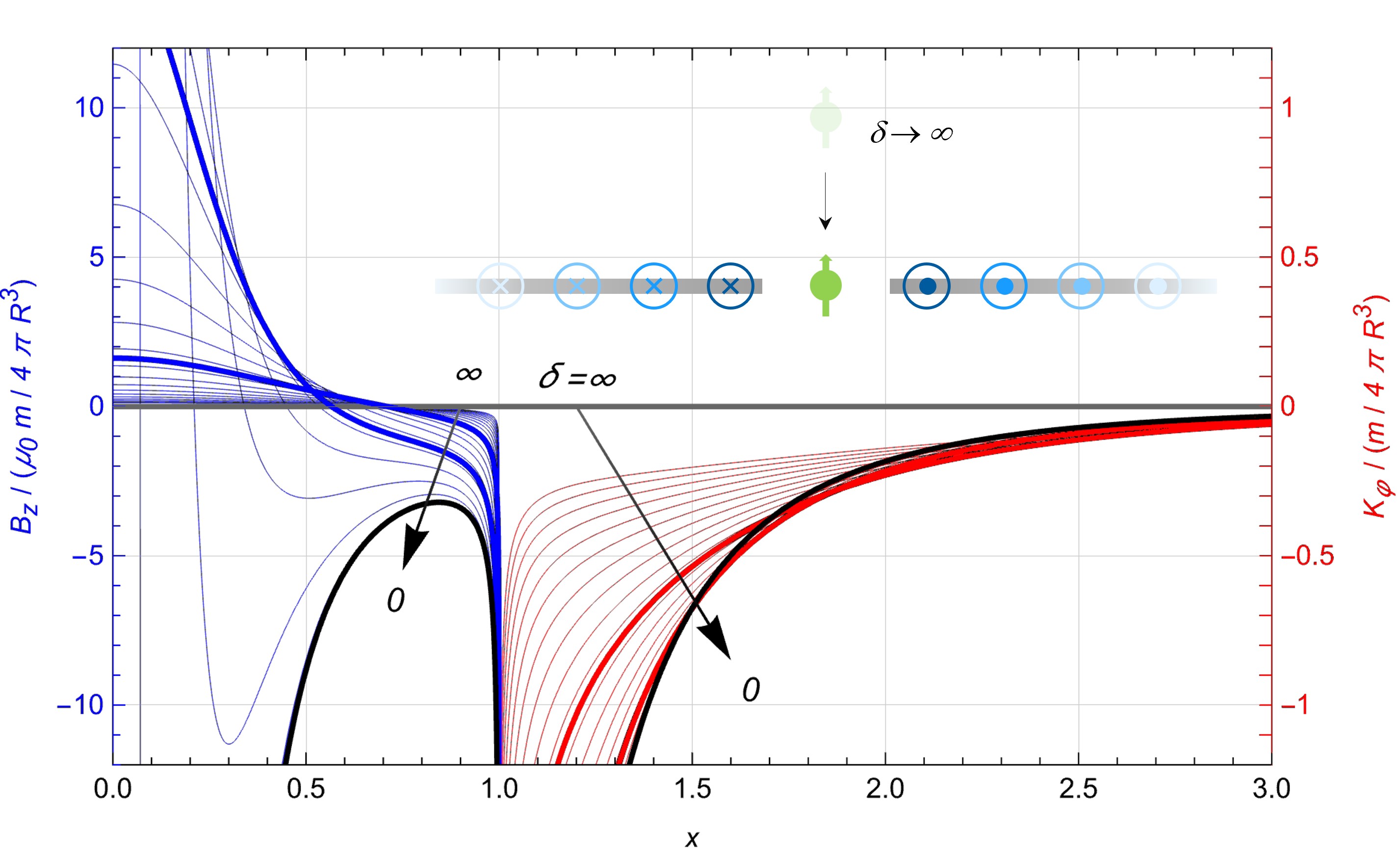}
	\caption{Current distribution $K_\varphi(x>1)$ and $B_z(x<1,0)$ component of the field at the hole when the superconductor has been zero-field cooled (currents density and fields are zero everywhere) and, after cooling, the dipole is moved along the $z-$axis. The plots are shown for $\delta=2$ to $\delta=0$ in intervals of $0.1$ following the arrows (the cases $\delta=0.5$ and $\delta=1$ are highlighted in thicker lines). The cases $\delta=0$ and $\delta\rightarrow \infty$ are highlighted in thick black and gray lines, respectively. The inset indicates that the system has been ZFC (dipole at $\delta\rightarrow \infty$ and moved towards $\delta=0$. The sketched currents are those at $\delta=0$}
	\label{fig:zfc}
\end{figure}

The surface current density and the $z$-component of the induction field in the hole can be written from Eqs. \eqref{eq:Kphidelta} and \eqref{eq:Bzdelta}, substituting $I\rightarrow I_{\rm ZFC}$. In Fig. \ref{fig:zfc}, these magnitudes for different values of $\delta$ are plotted. The case $\delta\rightarrow \infty$ gives zero current and zero field in the $z=0$ plane (marked in thick gray lines in Fig. \ref{fig:zfc}). When the dipole reaches the centre of the hole ($\delta=0$) the surface current density and the $z$ component of the induction field are
\begin{eqnarray}
    K_\varphi(x>1) &=& - \frac{m}{\pi^2 R^3} \frac{2}{x^3\sqrt{x^2-1}}, \\
    B_z(x<1,0) &=& B_z^{\rm a}(x,0)  \nonumber \\ &+& \frac{\mu_0 m}{2\pi^2 R^3}\Bigg[ -\frac{1+x^2}{x^2\sqrt{1-x^2} }   \nonumber \\ &+&\frac{1}{x^3} \tan^{-1}\left(\frac{x}{\sqrt{1-x^2}}\right)\Bigg].
\end{eqnarray}
These expressions are highlighted in Fig. \ref{fig:zfc} with thick black lines.

\subsection{Superconductor with a slit}
If one does not want to have flux restrictions, one can make a slit (from the hole to infinity) in the superconductor \cite{Brandt2005,Brandt2006,Navau2014}. The slit is assumed so narrow that its only effect is to prevent currents to circle the hole. In this case, $I=0$ is imposed. The flux threading the hole does not have to be kept when varying the field, since there is no way of having a closed line encircling the hole completely inside the superconductor. Thus, this would correspond to the $I=0$ case, independent of whether the superconductor has been field cooled or zero-field cooled.

\begin{figure}[t!]
	\centering
		\includegraphics[width=0.48\textwidth]{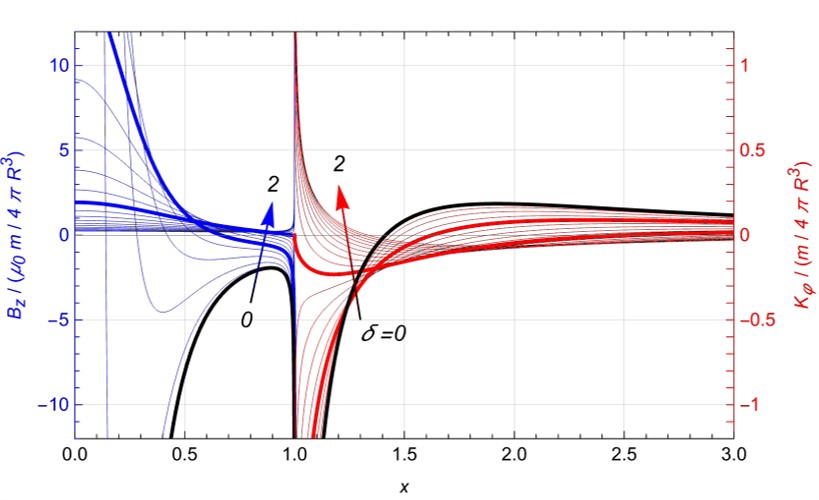}
	\caption{Current distribution $K_\varphi(x>1)$ and $B_z(x<1,0)$ component of the field at the hole for different $\delta$, when the superconductor has a narrow slit. The plots are shown for $\delta=0$ (highlighted in thick black line) to $\delta=2$ in intervals of $0.1$ following the arrows (the cases $\delta=0.5$ and $\delta=1$ are highlighted in thicker lines).}
	\label{fig:slit}
\end{figure}

In Fig. \ref{fig:slit} the surface current densities and $z$ component of the field are plotted as a function of the position of the dipole along the axis. The equations are the same as Eqs. \eqref{eq:KphiFC} and \eqref{eq:BzFC}. 

\begin{figure}[t!]
	\centering
		\includegraphics[width=0.47\textwidth]{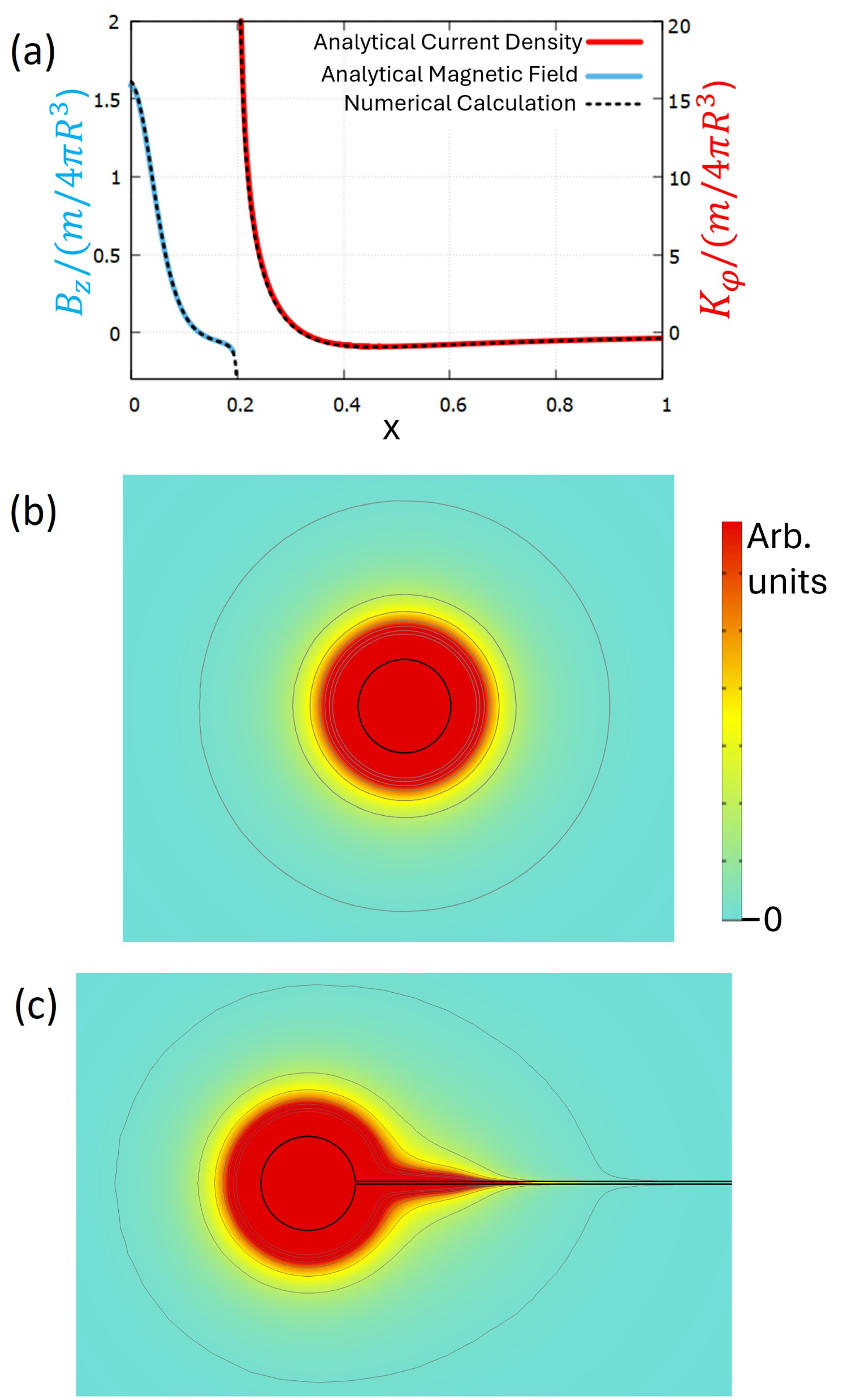}
	\caption{(a) Comparison of the numerical values obtained for the field in the hole and the currents in the holed superconductor with the analytical results for the case ($I=0$, $\delta=0.0125$) used as a double check. The match is perfect. (b) and (c): For $\delta=0.0125$, numerically evaluated field modulus at the $z\rightarrow 0^+$ for the case of a superconductor without a slit with $I=0$ (b) and with a slit (c). The colorscale is in arbitrary units, where red indicates a stronger field and blue a lower field. The thicker (black) lines indicate the hole and the slit and the thinner (gray) ones indicate the lines with the same modulus of the magnetic field.}
	\label{fig:fieldslit}
\end{figure}

Note that in this case, the slit would break the cylindrical symmetry and the found expressions would be only approximate for narrow slits. In Fig. \ref{fig:fieldslit}, we present numerically calculated values for the field just above the $z=0$ surface. In this figure, apart from double-checking our analytical equation in a particular case, it clearly shows that, as expected, the slit distorts the cylindrical symmetry of the entire system. Interestingly, in the region close to the hole, where the field is stronger, the distortion due to the slit is more concentrated very near the slit. Conversely, far from the hole, where the fields and induced currents are weaker, the distortion extends to areas further from the slit.

\subsection{On the finiteness of the superconductor and the value of $\lambda$}
We are considering that the superconductor is infinite in the $x$ and $y$ dimensions. Our situation is the limiting one of a ring of external radius much larger than the internal one, $R$. It is essential in our results and discussions that the external field is localized close to the axis of the hole. If a uniform field were externally applied, we would need to take into account the large currents that would be present in the external radius of the ring. Moreover, the fluxoid, not the flux threading the hole, would be the magnitude to be fixed when changing the externally applied field. In our case, the surface current density always decays to zero for $\rho\gg R$ due to the localized source of the applied field.
	
The total flux that crosses the whole plane $z=0$ due to a centered sphere alone (without a superconductor) is zero. When evaluating the total flux that threads through the hole when the superconductor with $I=0$ is present, we have found that it is $\Phi=\mu_0 m/\pi R$. This would mean that a total $-\mu_0 m/\pi R$ flux must return beyond the external border of the superconducting plate. When the plate becomes larger and larger, the magnetic field that should be present beyond the exterior border tends to zero, as expected, although the total flux crossing the exterior can be non-zero since the available exterior surface for crossing tends to infinity. There is no contradiction in this result. 

As for the value of $\lambda$, note that while thin superconducting with $\lambda=0$
completely shields the field $\B$ in the superconducting region, a $\lambda>0$ or $\lambda^2/t > \lambda$ will allow $\B$ to penetrate the film from the $\rho=R$ region to its inside \cite{Brandt2004}. Thus, as $\lambda$ increases, the effective radius of the hole would increase from $R$, decreasing the flux threading the hole (part of the flux would flow through the SC region) and also decreasing the circulating current. Actually, when $\lambda\longrightarrow\infty$, the superconductor becomes transparent to the magnetic field and no current is induced in the superconducting region.

\section{Conclusions}
\label{sec:conclusion}

In the present case, we have shown that when a magnetic dipole is in the presence of a superconducting plane with a hole, the interaction through the induced currents should be carefully analyzed, since different scenarios are possible. There is no freedom to choose all the physical magnitudes. In particular, we have demonstrated that different mathematical solutions appear, corresponding to different physical situations. All these behaviors can be understood through a simple equation that relates the flux threading the hole, the net current circling it, and the applied field. We have also seen that the physical scenarios correspond to fixing either the net current circling the hole or the flux threading it.

There are a few cases in which a problem can be solved analytically as the present cylindrically symmetric case. However, the importance of these cases goes far beyond being just "academic" solutions. They give us the principal trends and the major key dependencies on the parameters of the studied system.  Numerical results can be highly precise and necessary in many situations, such as non-axial dipoles, dipoles outside the axis, or non-circular holes. However, some general ideas would be still valid. In particular, a flux-current-field relation must exist, although the explicit simple equation was unknown.

The analytical results and equations developed here can thus help and guide current and ongoing experiments to obtain ultrasensitive magnetomechanical sensors with potential applications in magnetometry, gravimetry, and quantum experiments involving massive objects. Furthermore, the presented results serve as a fundamental benchmarking case against which more elaborate models, different geometries, experimental results, and/or other scenarios can be compared.

\begin{acknowledgments} 
We acknowledge financial support from the Spanish Ministry of Science, Innovation and Universities MCIU/AEI through PID2023-149054NB-I00 and the Horizon Europe 2021-2027 Framework Programme (European Union) through the SUPERMEQ project (Grant Agreement number 101080143). J.C.-S. acknowledges funding from AGAUR-FI Joan Or\'o grants (2024 FI-2 00143) of the Generalitat de Catalunya and the European Social Fund Plus.
\end{acknowledgments}

\bibliography{bib_holesc} 

\end{document}